\documentclass[useAMS,usenatbib]{mn2e}
\usepackage{amsmath}
\usepackage{graphicx}
\usepackage{txfonts}
\usepackage{natbib}

\title[]{The most plausible explanation of the cyclical period changes in close binaries: the case of the RS CVn-type
binary WW Dra}

\author[    `    \& Qian S.-B.]{Liao W.-P.$^{1,2,3}$\thanks{E-mail:
liaowp@ynao.ac.cn (LWP)}and Qian S.-B.$^{1,2,3}$\\
$^{1}$ National Astronomical Observatories/Yunnan Observatory,
Chinese Academy of Sciences, P. O. Box 110, 650011 Kunming, P. R.
China.\\
$^{2}$ Key Laboratory for the Structure and Evolution of Celestial
Objects,
Chinese Academy of Sciences.\\
$^{3}$ Graduate School of the CAS, 100049, Beijing, P.R. China/Yuquan Road 19\#
, SijingShan Block, 100049, Beijing City, P. R. China.}

\begin{document}

\pagerange{\pageref{firstpage}--\pageref{lastpage}}

\label{firstpage}

\maketitle

\begin{abstract}

Cyclical period changes are a fairly common phenomenon in close
binary systems and are usually explained as due to \textbf{either}
the magnetic activity of one or both components
\citep[e.g.,][]{app92} or to the light-travel time effect(LTTE) of a
third body. We searched the orbital period changes in 182 EA-type
(including the 101 Algol systems used by \cite{hal89}), 43 EB-type
and 53 EW-type binaries with known both the mass ratio and the
spectral type of their secondary components. We reproduced and
improved the same diagram as Hall's (1989) according to the new
collected data. Our plots do not support the conclusion derived by
\cite{hal89} that all cases of cyclical period changes are
restricted to binaries having the secondary component with spectral
types later than F5. The presence of period changes also among stars
with secondary component of early type indicates that the magnetic
activity is one cause, but not the only one, for the period
variation. It is discovered that cyclic period changes, likely due
to the presence of a third body are more frequent in EW-type
binaries among close binaries. Therefore, the most plausible
explanation of the cyclical period changes is the LTTE via the
presence of a third body. By using the century-long historical
record of the times of light minimum, we analyzed the cyclical
period change in the Algol binary WW Dra. It is found that the
orbital period of the binary shows a $\sim112.2
\textbf{\textrm{yr}}$ cyclic variation with an amplitude of
$\sim0.1977\textbf{\textrm{days}}$. The cyclic oscillation can be
attributed to the LTTE via a third body with a mass no less than
$6.43 M_{\odot}$. However, no spectral lines of the third body were
discovered indicating that it may be a candidate black hole. The
third body is orbiting the binary at a distance shorter than 14.4 AU
and it may play an important role in the evolution of this system.
\end{abstract}

\begin{keywords}
binaries : close -- binaries : eclipsing -- stars: individual: WW
Dra-- stars: late-type.
\end{keywords}

\section[]{Introduction}

Orbital period changes of stellar eclipsing binary systems can be
investigated by analyzing the $(O-C)$ diagram showing the difference
between the observed epochs of light minimum and those computed with
a given ephemeris. A periodic pattern in an $O-C$ curve is a fairly
common phenomenon in \textbf{Algols}, W Ursae Majoris binaries, and
RS Canum Venaticorum and the cataclysmic variables \cite[]{hal89,
hal80, hob94, war88}. Similar patterns of the $O-C$ diagram for
several classes of close binaries suggest a common underlying
mechanism \cite[]{zav02}, such as mass loss, apsidal motion,
magnetic activity, and \textbf{presence} of a third body.
\cite{zav02} thought that apsidal motion and mass loss are unlikely
mechanisms. Therefore, at present, cyclical period changes can
usually be explained as due to \textbf{either} the magnetic activity
of one or both components \citep[e.g.,][]{app92} or to the
light-travel time effect(LTTE) via the presence of a third body.

The hypothesis that cyclical period changes are caused by the
presence of a third body has been discussed by several investigators
\cite[]{fri73, cha92a, bor96}. In this hypothesis, the motion of the
binary around the center of mass of the system causes a periodic
change in the observed period due to a light-travel time
effect(LTTE), thereby creating a periodic pattern in $O-C$ curve.
Afterwards, Hall found a striking correlation between the spectral
type of the low-mass secondary component and the presence of a
cyclical period change in his study \cite[]{hal89} on 101 Algols.
From his plot, he noted that all cases of cyclical period changes
are restricted to systems with spectral types of the secondaries
later than F5. Based on this result, \cite{app92} and \cite{lan98}
developed a theory to explain the periodic pattern in $O-C$ curves
of these systems. In this \textbf{theory}, a certain amount of
angular momentum is periodically exchanged between the inner and the
outer parts of the convection zone, and therefore the rotational
oblateness of the star and hence the orbital period changes while
the system's component goes through its activity cycles. However,
the period changes of those Algols used by \cite{hal89} were mainly
derived from visual and photographic observations. \textbf{This
theory was frequently used to interpret the orbital period
modulation of close binaries containing at least one cool component
\citep[e.g.,][]{hal91, qia99, qia00}}. As new and more accurate
observational material has accumulated since then, \textbf{in} the
present work, we will reproduce and improve the same diagram as
Hall's (1989) of EA, EB and EW-type binaries based on the new
collected data. We will check Hall's plot and discuss the cause of
cyclical period changes. Meanwhile, we will analyze the cyclical
period change of the RS CVn-type binary WW Dra derived from the
century-long historical record of the times of light minimum and
discuss its plausible cause.

\section[]{The most plausible explanation of the cyclical period changes}

As discussed above, at present, the magnetic activity of one or both
components \citep[e.g.,][]{app92} or the light-travel time
effect(LTTE) via the presence of a third body are usually invoked to
explain \textbf{the cyclical} period changes of close binaries.
\cite{hal89} searched the orbital period changes of 101 Algol-type
binaries in \cite{giu83}. The samples of our study are made by the
stars listed in \cite{kre01}, the 101 Algol systems in \cite{giu83},
and the Algol-type binaries listed in \cite{iba06}. As selection
criterion we considered stars that either show cyclical period
changes or have secondary component of late spectral type. Finally,
182 EA-type (including the 101 Algol systems used by \cite{hal89}),
43 EB-type, and 53 EW-type binaries were collected for this study.

In this paper, the data of EA, EB, and EW-type binaries are
presented in Tables 1 - 3, respectively. In Table 1, Column (1) and
(7) give the systems we selected; (2) and (8) the secondary
component's spectral type; (3) and (9) the mass ratio; (4) and (10)
the form of the period change; (5) and (11) the geometrical
structure of the binary : semidetached binaries(SD) or detached
binaries(D), and (6) and (12) the reference for the $O-C$
information.

In Table 2 and 3, Column (1) gives the systems we selected; (2) the
secondary component's spectral type; (3) the mass ratio; (4) the
form of the period change and (5) the reference for the $O-C$
information. The secondary component's spectral type
($\textrm{Sp}_{2}$) and the mass ratio (\textit{q}) are up to date
values taken from one of these references: \cite{kre01},
\cite{giu83}, \cite{iba06}, the reference given in corresponding
table, and VizieR database \footnote{http://vizier.u-strasbg.fr/,
operated at CDS, Strasbourg, France.}. Therefore, some of the
secondary component's spectral types in Table 1 differ from those in
\cite{giu83}. Moreover, in the process of investigation, we
reclassified several systems as EB type or detached binaries with
respect to \cite{giu83} according to the mentioned more recent
bibliography.

The plots of mass ratio(\textit{q}) vs. secondary component's
spectral type ($\textrm{Sp}_{2}$) for EA, EB, and EW-type binaries
are displayed in Figs. 1 - 3, respectively. The form of the period
change follows the convention adopted by \cite{hal89}. A horizontal
line ( - ) indicates no period change, a forward slash ( / )
indicates a period increase only, a back slash ( $\setminus$ )
indicates a period decrease only, a cross ( $\times$ ) indicates
both increase and decrease of the period, and a filled circle (
$\bullet$ ) is used for systems for which we have inadequate data
for judgement. In Fig. 1, the magenta symbols are used for the
semidetached Algol-type binaries and the black ones are for detached
Algol-type binaries. It is clear from Fig. 1 that our plots do not
support the conclusion derived by \cite{hal89} that all cases of
cyclical period changes are restricted to binaries with secondary
component with spectral type later than F5. There are cases among
both semidetached and detached Algols in which the spectral type of
the secondary component is earlier than F5, such as RW Cap
\cite[]{erd07}, TX Her \cite[]{ak04}, and it is expected that the
number of these systems will grow rapidly as more new observational
data will be derived. The presence of period changes also among
systems with low-mass component of early-type stars \textbf{rules
the magnetic activity out as unique cause} for the period variation.
Among binaries with late-type component the orbital period variation
can be due to either magnetic activity or LTTE. Whereas, among
binaries with early-type components the LTTE is the more likely
cause. Moreover, the validity of the Applegate mechanism has
recently come into question \cite[]{lan05, lan06}. Lanza suggested
that the Applegate mechanism should be rejected because it can not
explain the orbital period modulations of classical RS CVn close
binaries \cite[]{lan05}. Afterwards he also found that the mechanism
is inadequate to explain the cyclical period changes of all close
binaries with a late-type secondary \cite[]{lan06}. Again, the
Applegate mechanism predicted that there is a connection between the
luminosity variation and the variation of period. However, to date,
no reliable connections were found in the literature. Therefore, the
most plausible explanation of the cyclical period changes is the
LTTE via the presence of a third body. We found that 48.9 \% of EA,
44.2 \% of EB and 64.2 \% of EW-type binaries have cyclical orbital
period variation. If we assume that such variations are related to
the presence of a third body through the LTTE, then we find that EW
stars have the highest probability to belong to multiple systems.
These results are in agreement with the findings of \cite{cha92b}.
The detailed statistical numbers of cyclical period changes in close
binary systems are displayed in Table 4.

In the following sections,we present our investigation on the
cyclical period change in the RS CVn-type binary WW Dra and discuss
about its causes as the presence of a black hole companion.

\begin{table*}
\tiny
\begin{center}
\caption{Period changes of semidetached and detached Algol-type
binaries.} \label{Table 1} \tiny
\begin{tabular}{lllccclllccc}
\hline\hline Star & $\textrm{Sp}_{2}$ & \textit{q} & Type of $\Delta$\textit{P}& Geo.Str& Ref.&Star & $\textrm{Sp}_{2}$ & \textit{q} & Type of $\Delta$\textit{P}& Geo.Str& Ref.\\
\hline
TT And     &   [G7IV]      &  0.29   &  $\times$    & SD&(1)&    RS Vul     &     G0III-IV& 0.310   &    -           &  SD& (35)\\
TW And     &    K0-K1      &  0.21   &  $\times$    & SD&(2)&    BE Vul     &    K2-K3    & 0.40    &    $\setminus$ &  SD& (17)\\
XZ And     &   G5IV        &  0.4    &  $\times$    & SD&(3)&    V78$\omega$Cen&  K2-K3   & 0.25    &    $\bullet$   &  SD& (36)\\
KO Aql     &  (F8IV)       & 0.223   &  /           & SD&(4)&    TY Del     &    G0IV     & 0.03    &    $\times$    &SD&(6, 10$^{+}$)\\
V342 Aql   &   [K0IV]      &  0.28   &  $\times$    & SD& (1)&   VX Lac     &    K4IV     & 0.32    &    $\times$    &  SD&  (6, 10$^{+}$)\\
V346 Aql   &   G           &  0.3    &  $\setminus$ & SD & (5)&  RV Per     &    [G74]    & 0.29    &    $\times$    &  SD& (6)\\
RY Aqr     &   G8?         &  0.230  &  $\times$    & SD& (6)&   BO Vul     &    G0IV     & 0.44    &    $\times$    &  SD& (6, 10$^{+}$)\\
SS Cam     &   F5V         &  0.954  &  $\times$    & SD& (7)&   IV Cas     &    G1V      & 0.5     &    $\times$    &  SD& (6, 10$^{+}$)\\
RW Cap     &   A4          &  0.450  &  $\times$    & SD& (1)&   BF CMi     &   [K0IV]    & 0.3     &    $\times$    &  SD& (6)\\
QZ Car     &   B0g         &  0.60   &  $\times$    & SD& (8)&   TY Cap     &   [G3.5IV]  & 0.4     &    $\times$    &  SD& (6)\\
AB Cas     &   K0          &  0.22   &  /           & SD & (5)&  DK Cep     &   [G4IV]    & 0.560   &    $\times$    & SD&  (6)\\
BZ Cas     &   [G1.5IV]    &  0.32   &  $\times$    & SD& (1)&   SS Cet     &   [KOIV]    & 0.25    &    $\times$    &  SD& (6)\\
RZ Cas     &    G5IV       &  0.351  &  $\times$    & SD& (9)&   RR Dra     &   [G8IV]    & 0.28    &    $\times$    &  SD& (6)\\
TV Cas     &    G2         &  0.470  &  $\times$    & SD& (10)&   UZ Sge     &   [G0IV]    & 0.14    &    $\times$    & SD&(6)\\
TW Cas     &   K4-K5       &  0.41   &  $\setminus$ & SD& (5, 11)& EW Lyr     &   [K3IV]    & 0.300   &    $\times$   & SD&(6)\\
U Cep      &    G8III-IV   &  0.550  &  $\times$     & SD&(12)&  YY Gem    &    M1V         & 1.006   &    $\times$  &  D  &(37)\\
XX Cep     &    (G4IV)     &  0.150  &  $\times$     & SD&(10)&  RX Her    &    A0V         & 0.847   &    -         &  D  &(5)\\
XY Cep     &   (G4IV)      &  0.250  &  $\setminus$  & SD& (13)&  TX Her    &    F2V         & 0.895   &    $\times$  &  D  &(38)\\
GT Cep     &    B9.5g      &  0.34   &  $\bullet$    & SD& (5)&  VZ Hya    &    F6V         & 0.911   &   $\bullet$  &  D  &(5)\\
R CMa      &    G8IV-V     &  0.170  &  $\times$     & SD& (6)&  HS Hya    &    F5V         & 0.971   &   $\bullet$  &  D  &(5)\\
TZ CrA     &   K2-K3       &  0.3    &  $\bullet$    & SD& (5)&  CM Lac    &    F0V         & 0.782   &   -          &  D  &(5)\\
U CrB      &    F8III-IV   &  0.289  &  $\times$     & SD& (3)&  UV Leo    &    G2V         & 0.917   &   $\times$   &  D  &(5)\\
RW CrB     &    K3         &  0.220  &  $\setminus$  & SD& (14)&  FL Lyr    &    G8V         & 0.787   &   -         &  D  &(5)\\
SW Cyg     &    K3         &  0.190  &  $\times$     & SD& (5)&  UX Men    &    F8V         & 0.968   &   $\bullet$  &  D  &(5)\\
UW Cyg     &   K4IV        &  0.28   &  $\times$     & SD& (15)& CD Tau    &    F7V         &  0.949  &   $\bullet$  &  D  &(5)\\
WW Cyg     &   (G9)        &  0.310  &  $\times$     & SD& (16)& DM Vir    &    F7V         &  0.991  &   $\bullet$  &  D  &(5)\\
ZZ Cyg     &    K6         &  0.52   &  $\setminus$  & SD & (5)& RS Ari     &    G5          & 0.360   &  $\bullet$  & D& (5)\\
KU Cyg     &    K5eIII     &  0.125  &  $\bullet$    & SD& (5)&  WW Aur     &    A7Vm        & 0.905   &    -           &  D & (5)\\
MR Cyg     &    B8         &  0.56   &  $\times$     & SD& (3)&  SS Boo     &    KIIV        & 0.988   &    $\bullet$   &  D& (5)\\
TT Del     &   G7IV        &  0.290  &  $\times$     & SD& (17)&  SV Cam    &    K4V         & 0.670   &    $\times$    &  D& (5)\\
W Del      &   K1-K2       &  0.18   &  $\times$     & SD& (3)&  RS CVn     &    F4V-IV      &  0.958  &    $\setminus$ &  D &(5)\\
Z Dra      &   K3-K4       &  0.23   &  $\times$     & SD& (12)& FZ CMa     &    B3IV-V      &  0.880  &    $\times$    &  D& (6, 53)\\
TW Dra     &   K0III       &  0.470  &  $\times$     & SD& (12)& RW UMa     &    K1IV        &  0.951  &    $\bullet$   &  D& (5)\\
AI Dra     &   GV-IV       &  0.429  &  $\times$     & SD& (18)&  BH Vir    &    G2V         &  0.950  &    $\bullet$   &  D& (5)\\
S Equ      &    K0-K1      &  0.131  &  $\times$     & SD& (12)&  HW Vir     &    M           &  0.2931 &    $\times$   &  D& (39, 40)\\
AS Eri     &    G6IV       &  0.110  &  $\bullet$    & SD& (5)&  XX Cas     &    B6n         &  0.36   &    $\bullet$   &  D& (5)\\
TZ Eri     &    K0-1III    &  0.19   &  $\times$     & SD& (6)&  AQ Cas     &    B9          & 0.28    &   $\bullet$    &  D& (5)\\
WX Eri     &   F6V:        &  0.29   &  $\bullet$    & SD& (5)&  GG Cas     &    K0III       & 0.78    &   $\bullet$    &  D& (5)\\
RW Gem     &   F5Ib-I      &  0.290  &  -            & SD &(5)& MN Cas     &    A0V         & 0.960   &   $\bullet$    &  D& (5)\\
AF Gem     &    G0III-IV   &  0.342  &  $\setminus$   & SD &(5)&ZZ Cep     &    F0V         & 0.460   &   $\bullet$    &  D& (5)\\
X Gru      &   K4-K5       &  0.27   &  /             & SD& (19)& UX Com     &    G2V         & 0.855   &   $\bullet$    &  D& (5)\\
$\mu$ Her  &   B6          &   0.36 &   -           & SD& (5)&   RT CrB     &    G0          & 0.991   &   $\times$     &  D& (41)\\
SZ Her     &   G8-G9       &  0.4    &  $\times$     &SD& (12, 20)&V909 Cyg   &  A2          & 0.850   &   $\bullet$    &  D& (5)\\
UX Her     &    K?         &  0.210  &  $\times$     & SD& (12, 21)& WW Dra   &   K0IV        & 0.985   &   $\times$     &  D& (42)\\
AD Her     &    K2         &  0.350  &  -            & SD& (22)& RZ Eri     &    K0          & 0.963   &   $\bullet$    &  D& (5)\\
V338 Her   &   K6          &  0.16   &  $\times$     & SD& (12)& V819 Her   &    F8V         & 0.704   &   $\times$     &  D& (43)\\
RX Hya     &   M5          &  0.24   &  $\times$     & SD& (15)& AW Her     &    K2          & 0.906   &   $\bullet$    &  D& (5)\\
TT Hya     &   G6III       &  0.224  &  $\bullet$    & SD& (5)&  DQ Her     &    M3V         & 0.66    &   $\times$     &  D& (44, 45$^{*}$)\\
TW Lac     &   [K0IV]      &  0.26   &  $\times$     & SD& (1)&  MM Her    &     G2V         & 0.944   &   $\bullet$    &  D& (5)\\
RW Leo     &   F4          &  0.24   &  $\times$     & SD& (2)&  AR Lac    &     K0IV        & 0.897   &   $\times$     &  D& (46)\\
Y Leo      &   K5          &  0.3    &  $\times$     & SD& (12)& AO Mon    &     B5          & 0.950   &   $\bullet$    &  D& (5)\\
RS Lep     &   M0          &  0.3    &  $\times$     & SD& (5)&  V635 Mon  &     A2          & 0.420   &   $\bullet$    &  D& (5)\\
$\delta$ Lib&  K2IV        &  0.345  &  /           &SD& (21, 23)&SZ Psc    &     F5-8V       & 0.766   &   $\bullet$    &  D& (5)\\
T Lmi      &    G5III      &  0.130  &  $\times$     & SD& (6)&  TY Pyx    &     G5IV        & 0.987   &   $\bullet$    &  D& (5)\\
TT Lyr     &   K0          &  0.27   &  -            & SD & (5)& AC Cnc    &     K2          & 0.77    &   $\times$     &  D& (47)\\
FW Mon     &   F2          &  0.36   &  $\bullet$    & SD& (5)&  PW Her    &     K0IV        & 0.768   &   $\times$     &  D& (41)\\
TU Mon     &   F3          &  0.210  &  -            & SD & (5)& EU Hya    &     G0          & 0.630   &   $\times$     &  D& (48)\\
AR Mon     &   K2III       &  0.30   &  $\bullet$    & SD& (5)&  QY Aql    &     K5          & 0.33    &   $\bullet$    &  D& (5)\\
RV Oph     &   K0          &  0.10   &  -            & SD & (12)& V805 Aql  &     (A9)        & 0.773   &   -            &  D &(5)\\
UU Oph     &   G8-G9       &  0.29   &  $\bullet$    & SD& (24)&  AR Aur    &     B9V         & 0.925   &   $\times$     &  D& (5)\\
AQ Peg     &   G5          &  0.21   &  $\times$     & SD& (12)&  $\beta$ Aur  &  A1IV        & 0.969   &    $\bullet$   &  D&(5)\\
AT Peg     &   K6          &  0.484  &  $\times$     & SD& (3)&  HS Aur    &     K0V         & 0.977   &   $\bullet$    &  D& (5)\\
DI Peg     &   K0          &  0.3    &  $\times$     & SD& (25)&  IM Aur    &     A5V:        & 0.431   &   $\times$     & D& (49)\\
RY Per     &   F7:2-3      &  0.271  &  -            & SD &(12)& SU Boo    &     K2          & 0.11    &   $\bullet$    &  D& (5)\\
RT Per     &   K2-K3       &  0.24   &  $\times$     & SD& (12)&  Y Cam     &     K0          & 0.240   &   $\times$     & D& (10)\\
ST Per     &   K1-2IV      &  0.150  &  $\times$     &  SD&(3)&  SZ Cam    &     B0.5V       & 0.25    &   $\times$     &  D& (50)\\
DM Per     &   A5III       &  0.284  &  $\times$     &  SD&(26)&  YZ Cas    &     F2V         & 0.584   &   $\bullet$    &  D& (5)\\
$\beta$ Per &  G8III        &  0.217 &   $\times$    &  SD&(27)&  SZ Cen    &     A7V         & 1.018   &   $\bullet$    &  D& (5)\\
Y Psc      &   K0IV        &  0.250  &  $\times$     &  SD&(5, 12)&  WX Cep    &  A2V         & 1.090   &   $\bullet$    &  D& (5)\\
XZ Pup     &   (K2IV)      &  0.400  &  $\bullet$    &  SD&(5)&  EI Cep    &     F1V         & 1.054   &   -            &  D & (5)\\
RZ Sct     &    A2IV       &  0.216  &  $\bullet$    &  SD&(5)&  RS Cep    &     G           & 0.145   &   /            &  D& (5)\\
EG Ser     &   A           &  0.905  &  $\bullet$    &  SD&(5)&  XY Cet    &     Am          & 0.926   &   $\bullet$    &  D& (5)\\
U Sge      &   G4III-IV    &  0.370  &  $\times$     &  SD&(28)&  S Cnc    &    G8-9III-IV  & 0.090    &   -            &  D & (5)\\
RS Sgr     &   A2V         &  0.36   &  $\bullet$    &  SD&(5)&  UZ Cyg    &     K1          & 0.07    &    -           &  D & (5)\\
XZ Sgr     &    G5IV-V     &  0.140  &  $\bullet$    &  SD&(5)&  VW Cyg    &     G5          & 0.280   &   /            &  D & (5)\\
V505 Sgr   &    F7IV       &  0.520  &  $\times$     &  SD&(29)&  BS Dra    &     F5V         & 1.000   &   -            &  D & (5)\\
AC Tau     &   [G]         & 0.683   & $\times$      &  SD&(15)&  CM Dra    &     M4V         & 0.873   &   $\bullet$    &  D& (5)\\
HU Tau     &    F5III-IV   & 0.256   & -             &  SD &(30)& AL Gem    &     K4          & 0.10    &    /           &  D & (5)\\
RW Tau     &   K0V         & 0.19    & $\times$      &  SD&(31)&  RX Gem    &     K2?         & 0.254   &    $\times$    &  D& (51)\\
$\lambda$ Tau&  A4IV       &  0.263  &   -           &  SD&(5)&  RY Gem    &     K0-3IV-V    & 0.193   &    $\setminus$ &  D & (5)\\
 X Tri      &    G3       & 0.51    &    $\times$    &  SD&(32)&  AU Mon    &    F8-G0II-III  & 0.199   &   $\bullet$  &  D& (5)\\
 V Tuc      &    (K2)     & 0.28    &    $\bullet$   &  SD&(5)&     FS Mon    &    F4V          & 0.896   &   $\bullet$ &  D& (5)\\
 TX UMa     &     F6IV    & 0.248   &    $\times$    &  SD&(12)&    U Oph     &     B6V         & 0.925   &   $\times$   &  D& (52)\\
 VV UMa     &    G6       & 0.21    &    $\times$    &  SD&(33)&    WZ Oph    &     F8V         &  0.982  &    -         &  D & (5)\\
 W Umi      &    (KOIV)   & 0.490   &    $\setminus$ &  SD&(5)&    DN Ori    &     G           &  0.07   &    $\bullet$ &  D& (5)\\
 RT UMi     &    K6       & 0.3     &    $\times$    &  SD&(5)&    AW Peg    &     F5IV        &  0.16   &    $\bullet$ &  D& (5)\\
 S Vel      &    K5IIIe   & 0.120   &    $\bullet$   &  SD&(34)&    BK Peg    &     F8V         &  1.117  &    $\bullet$ &  D& (5)\\
 DL Vir     &    K0IV     & 0.570   &    $\bullet$   &  SD&(5)&    EE Peg    &     F5V         &  0.619  &    $\bullet$ &  D& (5)\\
 UW Vir     &    K3IV     & 0.23    &    $\times$    &  SD&(15)&    RW Per    &    G0III        &  0.150  &    $\times$  &  D& (54)\\
 AY Vul     &    K        & 0.24    &    $\times$    &  SD&(2)&   $\varsigma$ Phe & B8V        &  0.651 &     $\times$  &  D&(43)\\
 Z Vul      &    A2III    & 0.430   &    $\times$    &  SD&(5)&    UV Psc    &    K3V          &  0.733  &    -         &  D & (5)\\

\hline
\end{tabular}
\end{center}
\tiny
Note. $^{+}$ the reference for the $\textrm{Sp}_{2}$. $^{*}$ the reference for the mass ratio q\\
\tiny References: (1) \cite{erd07}; (2) \cite{qia03a};
(3)\cite{bor96}; (4) \cite{hay79}; (5) \cite{kre01}; (6)
\cite{zas05}; (7) \cite{qia09a}; (8) \cite{may01}; (9) \cite{nar94};
(10)\cite{hof06}; (11) \cite{mcc71}; (12) \cite{kre71};
(13)\cite{kre88}; (14)\cite{qia00a}; (15)\cite{qia00b};
(16)\cite{zav02};
(17)\cite{qia01a};(18)\cite{deg00};(19)\cite{sch77};
(20)\cite{bal97}; (21)\cite{bal96}; (22)\cite{bat78};
(23)\cite{koc62}; (24)\cite{tay81}; (25)\cite{lu92};
(26)\cite{hil86}; (27)\cite{sod80}; (28)\cite{sim97};
(29)\cite{qia98}; (30)\cite{par80}; (31)\cite{fri73};
(32)\cite{rov00}; (33)\cite{sim96}; (34)\cite{sis71};
(35)\cite{kre78}; (36)\cite{sis69}; (37)\cite{qia09b};
(38)\cite{ak04}; (39)\cite{qia08a}; (40)\cite{lee09};
(41)\cite{qia03b}; (42)This paper ; (43)\cite{zas09};
(44)\cite{dai09}; (45)\cite{muk03}; (46)\cite{fra00};
(47)\cite{qia07a}; (48)\cite{qia03c}; (49)\cite{may90};
(50)\cite{lor98}; (51)\cite{qia02}; (52)\cite{wol02};
(53)\cite{mof83}; (54)\cite{may84};
\end{table*}

\begin{table}
\begin{minipage}{9cm}
\caption{Period changes of EB-type binaries.} \label{Table 2}
\begin{center}
\tiny
\begin{tabular}{lllcc}

\hline\hline Star & $\textrm{Sp}_{2}$ & \textit{q} & Type of $\Delta$\textit{P}& Ref.\\
\hline
AD And     &     A0V    &   1.000   &  $\times$    &  (1) \\
V337 Aql   &     B1.5   &   0.6     &  $\setminus$ &  (2)  \\
SX Aur     &     B3V    &   0.57345 &  /           &  (3, 4$^{*}$)\\
TT Aur     &     B4     &   0.668   &  $\times$    &  (5) \\
BF Aur     &     B5V    &   1.048   &  $\bullet$   &  (3, 6$^{*}$)\\
HL Aur     &     G9V    &   0.722   &  $\times$    &  (7) \\
IU Aur     &     B0.5V  &   0.672   &  $\times$    &  (8) \\
LY Aur     &     B0.5III&   0.62    &  $\bullet$   &  (3, 9$^{*}$)\\
YY CMi     &     F5     &   0.89    &  $\bullet$   &  (3, 10$^{*}$)\\
RX Cas     &     gG3    &   0.31    &  /           &  (3, 11$^{*}$)\\
CC Cas     &     O8     &   0.415   &  $\bullet$   &  (3, 12$^{*}$) \\
CR Cas     &     B1V    &   0.769   &  $\bullet$   &  (3, 13$^{*}$) \\
ZZ Cas     &     [B9]   &   0.7     &  $\times$    &  (14) \\
SV Cen     &     B6III  &   0.707   &  $\setminus$ &  (3, 15$^{*}$)\\
BH Cen     &     B3?    &   0.84    &  $\times$    &  (16) \\
EG Cep     &     F2V    &   0.464   &  $\times$    &  (17) \\
V701 Cen   &     A3IV-V &   0.617   &  $\bullet$   &  (18$^{+}$$^{*}$) \\
V758 Cen   &     A9     &   0.387   &  $\bullet$   &  (19$^{+}$$^{*}$) \\
AH Cep     &     B0.5V  &   0.87    &  $\times$    &  (20) \\
CQ Cep     &     O7     &   1.03    &  $\times$    &  (21, 22$^{*}$) \\
GK Cep     &     A0V    &   0.913   &  $\times$    &  (3, 23$^{*}$) \\
RV Crv     &     G0:    &   0.3     &  $\bullet$   &  (3, 24$^{*}$) \\
GO Cyg     &     A0n    &   0.428   &  /           &  (25 )  \\
KR Cyg     &     (F5)   &   2.04    &  $\bullet$   &  (3, 26$^{*}$) \\
V382 Cyg   &     O7.7V     &   0.551   &  $\times$    &  (27) \\
V448 Cyg   &     B1Ib   &   0.555   &  $\bullet$   &  (3) \\
V548 Cyg   &     F7     &   0.220   &  $\setminus$ &  (3)  \\
V729 Cyg   &     O8     &   0.282   &  $\bullet$   &  (3, 28$^{*}$) \\
RT Lac     &     G9IV   &   0.401   &  $\times$    &  (29, 30$^{*}$) \\
XX Leo     &     [F2]   &   0.82    &  $\times$    &  (31) \\
$\beta$ Lyr      &     A5     &   0.223   &  /     &  (3, 32$^{*}$) \\
TU Mus     &     O9.5V  &   0.651   &  $\times$    &  (27) \\
V Pup      &     B3     &   0.55    &  $\times$    &  (33, 34$^{*}$) \\
UZ Pup     &     A5     &   0.80    &  $\bullet$   &  (3, 35$^{*}$) \\
RT Scl     &     F0     &   0.443   &  $\setminus$ &  (3, 36$^{*}$)  \\
RY Sct     &     B0     &   0.51    &  $\bullet$   &  (3) \\
RU UMi     &     K      &   0.327   &  $\bullet$   &  (3) \\
AC Vel     &     B7     &   1.00    &  $\bullet$   &  (3, 37$^{*}$) \\
BF Vir     &     G2     &   0.331   &  $\setminus$ &  (3, 38$^{*}$)  \\
AS Ser     &     M0     &   0.3     &  $\times$    &  (39) \\
GW Tau     &     [A8]   &   0.309   &  $\times$    &  (40) \\
V701 Sco   &     B1-1.5 &   1       &  $\times$    &  (16) \\
IR Cas     &     [F9IV] &   0.2     &  $\times$    &  (41) \\
\hline
\end{tabular}
\end{center}
\end{minipage}
\tiny
Note. $^{+}$ the reference for the $\textrm{Sp}_{2}$. $^{*}$ the reference for the mass ratio q\\
\tiny
 References: (1) \cite{lia09a}; (2) \cite{may87}; (3) \cite{kre01}; (4) \cite{lin88}; (5) \cite{Ozd01}; (6) \cite{kal00};
 (7) \cite{qia06a}; (8) \cite{Ozd03}; (9) \cite{dre89}; (10) \cite{viv99}; (11) \cite{tod93}; (12) \cite{hil94}; (13) \cite{vit07};
 (14) \cite{lia09b}; (15) \cite{ruc92}; (16) \cite{qia06b}; (17) \cite{zhu09}; (18) \cite{mil88}; (19) \cite{lip85}; (20) \cite{kim05};
 (21) \cite{bor96}; (22) \cite{dem97}; (23) \cite{pri09}; (24) \cite{giu82}; (25) \cite{zab06}; (26) \cite{al85};
 (27) \cite{qia07b}; (28) \cite{rau99}; (29) \cite{iba01}; (30) \cite{cak03}; (31) \cite{zas05}; (32) \cite{lin98};
 (33) \cite{qia08b}; (34) \cite{sti98}; (35) \cite{mal89}; (36) \cite{ban90}; (37) \cite{joh97}; (38) \cite{rus81};
 (39) \cite{zhu08}; (40) \cite{zhu06}; (41) \cite{zhu04};

\end{table}

\begin{table}
\begin{minipage}{9cm}
\caption{Period changes of EW-type binaries.} \label{Table 3}
\begin{center}
\tiny
\begin{tabular}{lllcc}

\hline\hline Star & $\textrm{Sp}_{2}$ & \textit{q} & Type of $\Delta$\textit{P}& Ref.\\
\hline
AB And     & G5      &  0.560  &     $\times$  &  (1, 2$^{*}$, 3) \\
AK Her     & F6V    &  0.277  &     $\times$  &  (1, 3)  \\
GZ And     & G5V    &  0.514  &     $\times$  &  (4, 5$^{*}$)  \\
S Ant      & F4    &  0.33   &     /  &   (6, 7$^{*}$) \\
V417 Aql   & F9V    &  0.362  &     $\times$  &  (8) \\
V803 Aql   & K4      &  1.0    &     $\bullet$  & (6, 9$^{*}$) \\
ZZ Boo     & F2V  &  0.969  &     $\bullet$  &  (6) \\
44i Boo    & G1V    &  0.5    &     /  &   (6, 10$^{*}$) \\
TX Cnc     & F8V    &  0.455  &     $\times$  &  (11) \\
RV CVn     & G1      &  0.820  &     $\bullet$  &  (6)  \\
VW Cep     & K0V    &  0.35   &     $\times$  &  (12) \\
EM Cep     & B1V  &  0.520  &    $\setminus $ &   (6) \\
TW Cet     & G5      &  0.530  &     $\setminus $  &   (6, 13$^{*}$) \\
AA Cet     & F2V    &  0.240  &     $\bullet$  &  (6)  \\
RW Com     & G2e    &  0.345  &     $\times$  &  (13$^{*}$, 14) \\
EK Com     & G9V    &  0.304  &     $\bullet$  &  (6, 13$^{*}$) \\
V865 Cyg   & G        &  0.446  &     $\bullet$  &  (6, 13$^{*}$) \\
BV Dra     & F8V    &  0.402  &     $\times$  &  (13$^{*}$, 15) \\
BW Dra     & G0V    &  0.280  &     $\times$  &  (6, 13$^{*}$) \\
YY Eri     & G5      &  0.400  &     $\times$  &  (13$^{*}$, 6) \\
WY Hya     & A6      &  0.850  &     $\bullet$  &  (6) \\
SW Lac     & G8p     &  0.797  &     $\times$  &  (16, 13$^{*}$, 6)\\
XY Leo     & K0V    &  0.500  &     $\times$  &  (6) \\
UV Lyn     & G0V     &  0.367  &     $\bullet$  &  (6, 13$^{*}$) \\
V502 Oph   & F9V    &  0.371  &     $\setminus $  &   (6, 13$^{*}$) \\
V566 Oph   & F4V    &  0.241  &     $\times$  &  (17 )  \\
V1010 Oph  & (F6)   &  0.340  &     $\setminus $  &   (6) \\
U Peg      & F3      &  0.315  &     $\setminus $  &   (6, 13$^{*}$) \\
Y  Sex     & [F8.5]  &  0.449  &     $\times$  &  (18) \\
V743 Sgr   & G8      &  0.319  &     $\bullet$  &  (6, 13$^{*}$) \\
RZ Tau     & A8V    &  0.540  &     $\times$  &  (13$^{*}$, 17) \\
W UMa      & F8V    &  0.488  &     $\times$  &  (6, 13$^{*}$) \\
BM UMa     & K        &  0.540  &     $\bullet$&  (6, 13$^{*}$) \\
AH Vir     & K0      &  0.420  &         /     &   (6, 13$^{*}$) \\
ER Vul     & G5V    &  0.957  &     $\bullet$  &  (6)  \\
BI CVn     & [F9]    &  0.50   &     $\times$  &  (19) \\
RZ Com     & G9      &  0.430  &     $\times$  &  (20) \\
VZ Lib     & [F1.5]  &  0.460  &     $\times$  &  (21) \\
AP Leo     & [GO]    &  0.460  &     $\times$  &  (22, 13$^{*}$)  \\
AD Cnc     & [K0]    &  0.620  &     $\times$  &  (23) \\
UX Eri     & [F9]    &  0.440  &     $\times$  &  (24, 13$^{*}$) \\
EQ Tau     & [G2]    &  0.442  &     $\times$  &  (25, 13$^{*}$) \\
AH Cnc     & [F5] &  0.5    &     $\times$  &  (26) \\
V899 Her   & [G]?    &  0.566  &     $\times$  &  (27) \\
IK Per     & [A2]    &  0.88   &     $\times$  &  (28) \\
TV Mus     & [F9.5]  &  0.25   &     $\times$  &  (29) \\
FG Hya     & [F8]    &  0.420  &     $\times$  &  (30)  \\
AO Cam     & [F8]    &  0.6    &     $\times$  &  (31) \\
AM Leo     & [F7.5] &  0.36   &     $\times$  &  (31) \\
RR Cen     & [F1.5] &  0.18   &     $\times$  &  (32) \\
EZ Hya     & [F8]   &  0.350  &     $\times$  &  (33)  \\
GR Vir     & [G1]    &  0.460  &     $\times$  &  (34) \\
AG Vir     & [F2]   &  0.160  &     $\times$  &  (17) \\
\hline
\end{tabular}
\end{center}
\end{minipage}
\tiny
Note.  $^{*}$ the reference for the mass ratio q\\
\tiny
 References: (1) \cite{bor96}; (2) \cite{pyc04};  (3)\cite{hof06}; (4) \cite{cha92a}; (5) \cite{dan06}; (6) \cite{kre01};
(7) \cite{due07}; (8) \cite{qia03b}; (9) \cite{liu08}; (10)
\cite{aln89}; (11) \cite{liu07}; (12) \cite{pri00}; (13)
\cite{pri03}; (14) \cite{qia02a}; (15) \cite{yan09}; (16)
\cite{pri99}; (17) \cite{qia01b}; (18) \cite{he07}; (19)
\cite{qia08c}; (20) \cite{he08}; (21) \cite{qia08d}; (22)
\cite{qia07c}; (23) \cite{qia07d}; (24) \cite{qia07e}; (25)
\cite{yua07}; (26) \cite{qia06c}; (27) \cite{qia06d}; (28)
\cite{zhu05}; (29) \cite{qia05a}; (30) \cite{qia05b}; (31)
\cite{qia05c}; (32) \cite{yan05}; (32) \cite{yan04}; (34)
\cite{qia04};
\end{table}

\begin{table}
\caption{Statistical numbers of cyclical period changes in close
binary systems.} \label{Table 4}
\begin{center}
\scriptsize
\begin{tabular}{lcccc}\hline\hline
&EA-type &EB-type & EW-type \\\hline
(1)Total No.& 106 (SD) + 76 (D) &43&53\\
(2)No. of $\times$& 66 (SD) + 22 (D) &19&34\\
(3)Binaries with $\textrm{Sp}_{2}$ earlier than F5 in (2)& 6 (SD) + 7 (D) &15&6&     \\
(4)ratio$^{+}$& 48.9\% &44.2\%&64.2\%\\
(5)ratio$^{*}$& 14.8\% &78.9\%&17.6\%\\
\hline
\end{tabular}
\end{center}
\tiny
Note. $^{+}$ the ratio of binaries show cyclical period changes to the total number.\\
      $^{*}$ the ratio of binaries not only show cyclical period changes but also have $\textrm{Sp}_{2}$ earlier than F5 to
      binaries show cyclical period changes.\\
\end{table}

\begin{figure}
\centering
\includegraphics[width=9.0cm]{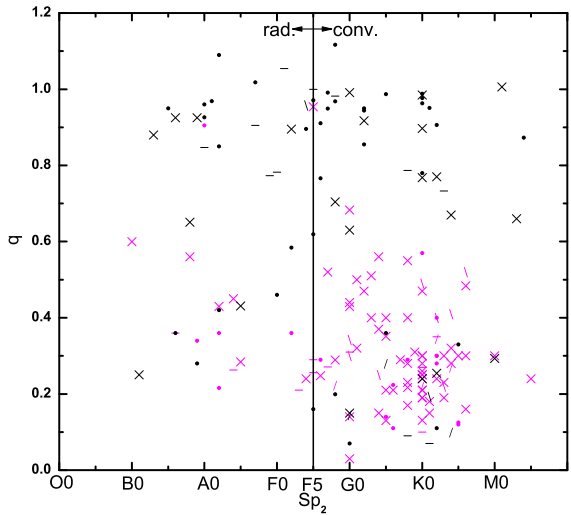}
\caption{Plot of mass ratio(\textit{q}) vs. secondary component's
spectral type ($\textrm{Sp}_{2}$) for EA-type binaries listed in
Table 1. A horizontal line ( - ) indicates no period change, a
forward slash ( / ) indicates a period increase only, a back slash (
$\setminus$ ) indicates a period decrease only, a cross ( $\times$ )
indicates both increase and decrease of the period, and a filled
circle ( $\bullet$ ) is used for systems for which have inadequate
data for judgement. The magenta symbols are for the semidetached
Algol-type binaries and the black ones are for detached Algol-type
binaries.} \label{Fig. 1}
\end{figure}

\begin{figure}
\centering
\includegraphics[width=9.0cm]{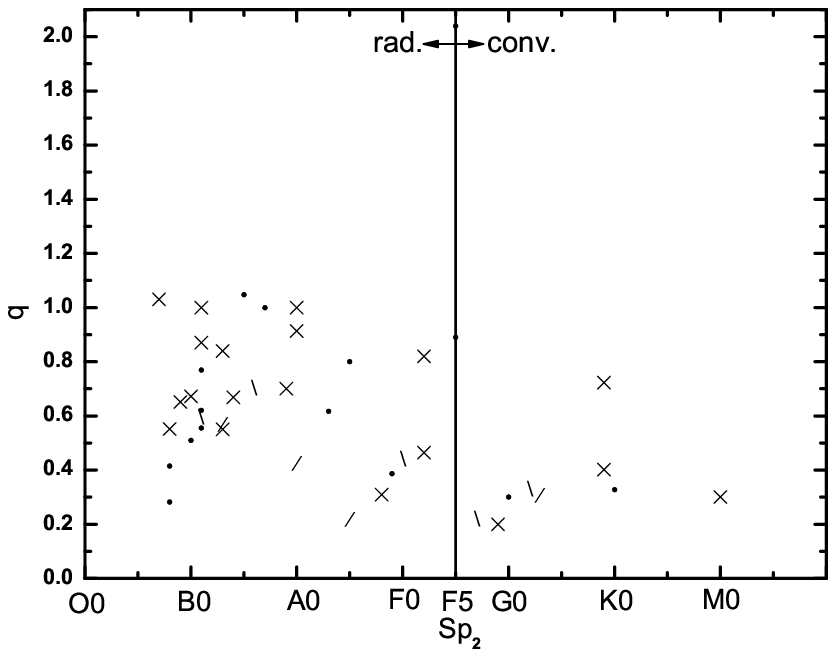}
\caption{Plot of mass ratio(\textit{q}) vs. secondary component's
spectral type ($\textrm{Sp}_{2}$) for EB-type binaries listed in
Table 2. The symbols are the same as in Figure 1.} \label{Fig. 2}
\end{figure}

\begin{figure}
\centering
\includegraphics[width=9.0cm]{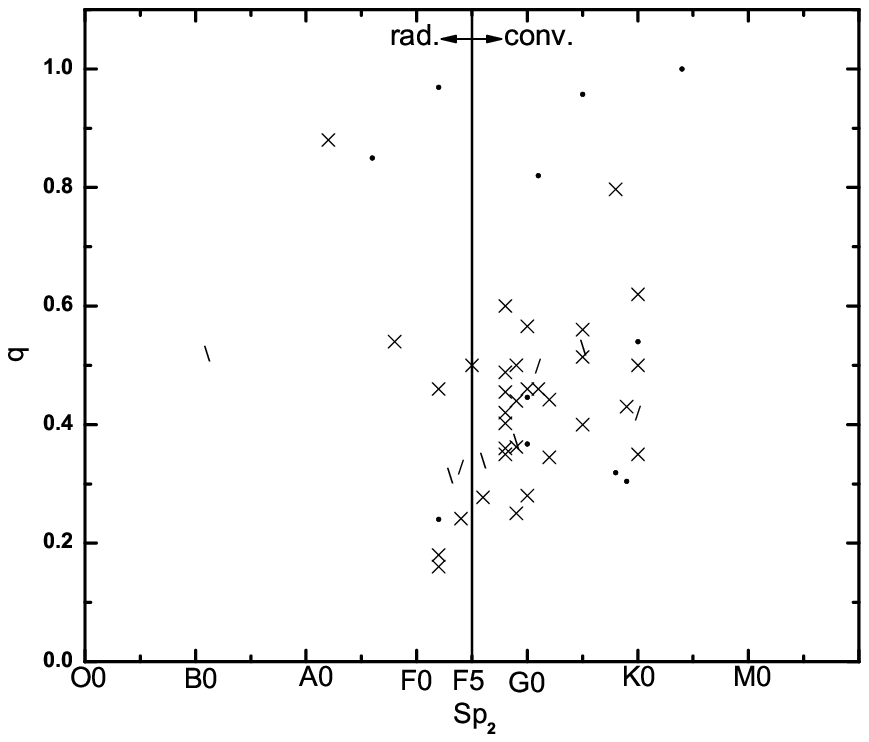}
\caption{Plot of mass ratio(\textit{q}) vs. secondary component's
spectral type ($\textrm{Sp}_{2}$) for EW-type binaries listed in
Table 3. The symbols are the same as in Figure 1.} \label{Fig. 3}
\end{figure}

\section[]{New CCD photometric observations for WW Dra}

WW Dra ( = HD 150708 = HIP 81519 = BD $+ 60\,^{\circ} 1691$,
$V_{max}=8.3\,mag$) was discovered to be an eclipsing binary by
\cite{har16}. \textbf{It is a RS CVn-type eclipsing binary with
G2+K0 spectral type \cite[]{joy41}.} Studies based on photographic
and photoelectric observations were carried out by \cite{pla40},
\cite{mez79}, \cite{mar80} and \cite{tun81}. Some of them also
calculated the orbital and physical elements of WW Dra, and the
binary was confirmed to be a detached system composed of two
sub-giant stars. The period variation of this binary was studied by
\cite{alb99} who derived the parameters of the light time orbit.
However, as more new observational data have been derived since
then, we will display different results of orbital period change
\textbf{of} WW Dra.

In order to analyze the period variations of WW Dra and investigate
the physical properties of the third body, CCD observations were
acquired on May 31, 2007 with the PI1024 TKB CCD photometric system
attached to the 1.0-m reflector at the Yunnan Observatory. The V
filter, close to the standard Johnson UBV system, was used. The
effective field of view is about $6^{\prime}.5\times 6^{\prime}.5$
at the Cassegrain focus and the size of each pixel is
$0^{\prime\prime}.38$. The integration time is 60 s for each image.
The coordinates of the nearby comparison star are RA=16:39:03.91,
DEC=+60:42:02.6 (J2000.0). The PHOT task of IRAF, which measures the
aperture magnitude for a list of stars, was used to reduce the
observed images. By using our photometric data, \textbf{we provided
the most recent determination of time of light minimum -} HJD
2454252.3029($\pm0.0018$).

\section[]{Orbital period variation of WW Dra}

To investigate the physical properties of the third body in WW Dra,
we search for cyclic orbital period changes. A total of 92 times of
light minimum from the literature have been collected and compiled
in the present paper. Most of times of light minimum were retrieved
from compilation of \cite{hal80}. Times of light minimum are listed
in the first and eighth column of Table 5. In our analysis, the
value HJD2443221.544 obtained by \cite{bud77} was not used because
its $(O-C)$ value shows large scatter when compared with the general
trend formed by the other data points. In the second and ninth
column the number (E) of orbits elapsed from the initial epoch of
primary minimum is listed. The $(O-C)_{1}$ values of all times of
light minimum were computed with the linear ephemeris given by
\cite{kre01},

\begin{eqnarray}
 Min.I=HJD\; 2427983.3236 + 4^{d}.6296328 E ,
\end{eqnarray}
\noindent \textbf{They are listed in the third and tenth column of
Table 5 and plotted vs. number of orbital periods in the} upper
panel of Fig. 4, where open circles refer to photographic or visual
observations, filled circles to CCD or photoelectric ones. As shown
in the upper panel of Fig. 4, the general $(O-C)_1$ trend can be
described by a linear curve with superimposed a periodic
fluctuation. Therefore, a sinusoidal term was added to a linear
ephemeris to get a good fit to the $(O-C)_1$ curve (solid line in
the upper panel of Fig. 4). \textbf{To} obtain a more accurate
result, we focus the fit to only primary minima, though
\textbf{secondary minima also} follow a similar trend. Weight 0.1
and 0.8 were assigned to lower-precision observations (photographic
or visual ones) and high-precision observations (CCD or
photoelectric ones), respectively. A weighted least-squares solution
yields the following equation,

\begin{eqnarray}
Min.I=2427983.2814(\pm0.0188)
       +4.^{d}6296389(\pm0.0000035)\times{E}\nonumber\\
+0.1977(\pm0.0096)\sin[0.^{\circ}0407\times{E}+163.^{\circ}73(\pm0.^{\circ}05)].
\end{eqnarray}
\noindent The sinusoidal term in Eq. (2) suggests a periodic
variation with a period of about 112.2 \textbf{\textrm{yr}} and an
amplitude of about $\textit{A} = {0^{d}.1977}$, which is more easily
seen from the lower panel of Fig. 4, where the linear part of Eq.
(2) was subtracted to the $(O-C)_1$ values. The good fit in Fig. 4
indicates no long-term steadily period increase or decrease.
Therefore, we can exclude the presence of mass transfer, which is in
accordance with the fact that WW Dra is a detached binary. The
$(O-C)_2$ values are shown in the fourth and eleventh column of
Table 5. The residuals of the fit with Eq. (2) are displayed in Fig.
5 and listed in the fifth and twelfth column of Table 5. \textbf{To}
detect possible regular trends in the residuals plotted in Fig. 5,
more high-precision times of light minimum are needed from future
observations.

\begin{table*}
\tiny
\begin{center}
\caption{$(O-C)$ data for WW Draco.} \label{Table 5}
\begin{tabular}{lcccccclcccccc}
\hline\hline JD.Hel. & E & $(O-C)_1$ & $(O-C)_2$ & Residuals & Weight &Ref.&JD.Hel. & E & $(O-C)_1$ & $(O-C)_2$ & Residuals & Weight &Ref.\\
2400000+ & & & & & & &2400000+ & & & & & &\\
\hline
15205.599   &-27607   &  0.0619  & 0.27336 &  0.11874    &   0.1   &(1)&39029.556    &2386     & -0.0715  & -0.1756 & 0.01954  &   0.1   &(1)\\
15501.809   &-2696    &  -0.0246 & 0.18294 &  0.02288    &   0.1   &(1)&39205.511    &2424     & -0.0425  & -0.14893& 0.04699  &   0.1   &(1)\\
15774.848   &-2637    &  -0.1339 & 0.07002 &  -0.09475   &   0.1   &(1)&40844.335    &2778     & -0.1085  & -0.23664& -0.04031 &   0.1   &(5)\\
15955.582   &-2598    &  0.0444  & 0.24593 &  0.07819    &   0.1   &(1)&41043.396    &2821     & -0.1217  & -0.25248& -0.05694 &   0.1   &(6)\\
16390.638   &-2504    & -0.0851  & 0.11066 &  -0.06367   &   0.1   &(1)&41154.547    &2845     & -0.0819  & -0.21415& -0.01914 &   0.1   &(7)\\
16418.486   &-2498    & -0.0148  & 0.18059 &  0.00585    &   0.1   &(1)&41168.406    &2848     & -0.1118  & -0.24424& -0.04929 &   0.1   &(7)\\
16603.702   &-2458    & 0.0158   & 0.20874 &  0.03144    &   0.1   &(1)&41682.288    &2959     & -0.1191  & -0.25834& -0.0666  &   0.1   &(1)\\
16969.44    &-2379    & 0.0128   & 0.2009  &  0.01897    &   0.1   &(1)&41682.298    &2959     & -0.1091  & -0.24834&  -0.0566 &   0.1   &(1)\\
17131.500   &-2344    & 0.0357   & 0.22165 &  0.03786    &   0.1   &(1)&41763.327    &2976.5   & -0.0986  &   -&       -  &   not used &(8)\\
18867.661   &-1969    & 0.0844   & 0.24735 &  0.05087    &   0.1   &(1)&41830.474    &2991     & -0.0813  & -0.22251& -0.0319  &   0.1   &(1)\\
18955.495   &-1950    & -0.0446  & 0.11719 &  -0.07956   &   0.1   &(1)&41904.533    &3007     & -0.0964  & -0.23859& -0.04859 &   0.1   &(9)\\
19191.586   &-1899    & -0.0649  & 0.09376 &  -0.10356   &   0.1   &(1)&41904.536    &3007     & -0.0934  & -0.23559& -0.04559 &   0.1   &(9)\\
19890.777   &-1748    & 0.0515   & 0.2009  &  0.00341    &   0.1   &(1)&41918.503    &3010     & -0.0153  & -0.15767&  0.03219  &   0.8   &(10)\\
22881.486   &-1102    & 0.0177   & 0.12749 &  -0.04559   &   0.1   &(1)&42617.379    &3161     & -0.2139  & -0.36553& -0.18264 &   0.1   &(11)\\
22895.458   &-1099    & 0.1008   & 0.2104  &  0.03752    &   0.1   &(1)&42904.6231   &3223     & -0.0070  & -0.16243&  0.01697  &   0.8   &(12)\\
25247.247   &-591     & 0.0364   & 0.11485    &  -0.01305 &  0.1    &(1)&42955.456   &3234     & -0.1001  & -0.25621& -0.07745 &   0.1   &(1)\\
27284.275   &-151     & 0.0260   & 0.07747    &  0.00209  &  0.1    &(1)&43043.489   &3253     & -0.0301  & -0.18737& -0.00977 &   0.1   &(1)\\
27284.32    &-151     & 0.0709   & 0.12237    &  0.04699  &  0.1    &(2)&43057.404   &3256     & -0.0040  & -0.16146& 0.01595  &   0.1   &(1)\\
27307.396   &-146     & -0.0012  & 0.04996    &  -0.02476 &  0.1    &(1)&43071.241   &3259     & -0.0559  & -0.21354& -0.03631 &   0.1   &(1)\\
27321.296   &-143     & 0.0098   & 0.06078    &  -0.01355 &  0.1    &(3)&43161.535   &3278.5   & -0.0397  & -&  - &   not used   &(1)\\
27335.197   &-140     & 0.0220   & 0.07279    &  -0.00115 &  0.1    &(1)&43189.321   &3284.5   & -0.0315  & - &  - &   not used   &(1)\\
27344.447   &-138     & 0.0127   & 0.06337    &  -0.01031 &  0.1    &(4)&43212.490   &3289.5   & -0.0107  & -&  -  &   not used   &(1)\\
27534.261   &-97      & 0.0118   & 0.05995    &  -0.00835 &  0.1    &(1)&43228.673   &3293     & -0.0314  & -0.19113& -0.01606&   0.1   &(1)\\
27543.530   &-95      & 0.0215   & 0.06953    &  0.00148  &  0.1    &(1)&43307.414   &3310     & 0.0058   & -0.15497& 0.01896 &   0.1   &(1)\\
27557.409   &-92      & 0.0116   & 0.05945    &  -0.0082  &  0.1    &(1)&43330.558   &3315     & 0.0017   & -0.15937& 0.01422 &   0.1   &(1)\\
27557.418   &-92      & 0.0206   & 0.06845    &  0.00079  &  0.1    &(1)&43344.442   &3318     & -0.0032  & -0.16446& 0.00894 &   0.1   &(1)\\
27645.3748  &-73      & 0.0144   & 0.06108    &  -0.00405 &  0.1    &(1)&43397.640   &3329.5   & -0.0460  & -&  - &   not used   &(1)\\
27654.6307  &-71      & 0.0110   & 0.05756    &  -0.00731 &  0.1    &(1)&43793.443   &3415     & -0.0766  & -0.24381& -0.07735 &   0.1   &(1)\\
27691.6670  &-63      & 0.0103   & 0.05637    &  -0.00744 &  0.1    &(1)&44168.521   &3496     & 0.0011   & -0.17107& -0.01102 &   0.1   &(13)\\
27710.1918  &-59      & 0.0165   & 0.06232    &  -0.00095  &  0.1    &(1)&44446.3404  &3556     & 0.0426   & -0.13325&  0.0217   &   0.8   &(14)\\
27881.4789  &-22      & 0.0072   & 0.05075    &  -0.00758 &  0.1    &(1)&44446.3406  &3556     & 0.0428   & -0.13305&  0.0219   &   0.8   &(15)\\
27904.6277  &-17      & 0.0079   & 0.05115    &  -0.00652 &  0.1    &(1)&44446.3408  &3556     & 0.0430   & -0.13285&  0.0221   &   0.8   &(14)\\
27918.5176  &-14      & 0.0089   & 0.05196    &  -0.0053  &  0.1    &(1)&44874.373   &3648.5   & -0.1659  & -&  - &   not used   &(13)\\
27918.5197  &-14      & 0.011    & 0.05406    &  -0.0032  &  0.1    &(1)&45284.239   &3737     & -0.0224  & -0.20935&  -0.0714 &   0.1   &(16)\\
27932.4068  &-11      & 0.0092   & 0.05208    &  -0.00478 &  0.1    &(1)&46175.376   &3929.5   & -0.0897  & -&  - &   not used   &(17)\\
27955.5536  &-6       & 0.0078   & 0.05037    &  -0.00581 &  0.1    &(1)&47631.6089  &4244     & 0.1237   & -0.0943&  -0.01512 &   0.8   &(18)\\
27983.3329  &0        & 0.0093   & 0.05151    &  -0.00387 &  0.1    &(1)&49534.4580  &4655     & 0.1937   & -0.04955&  -0.02578 &   0.8   &(19)\\
28020.3700  &8        & 0.0093   & 0.05102    &  -0.00329 &  0.1    &(1)&51636.31    &5109     & 0.1924   & -0.07869&  -0.11832 &   0.1   &(20)\\
28057.4074  &16       & 0.0097   & 0.05092    &  -0.0023  &  0.1    &(1)&52217.314   &5234.5   & 0.1775   & -&  - &   not used   &(21)\\
28205.5510  &48       & 0.0050   & 0.04426    &  -0.00462 &  0.1    &(1)&52416.431   &5277.5   & 0.2203   & -&  - &   not used   &(21)\\
28219.4445  &51       & 0.0096   & 0.04868    &  0.0002   &  0.1    &(1)&52576.270   &5312     & 0.3370   & 0.05345  &  -0.01358 &  0.1   &(22)\\
28307.4054  &70       & 0.0075   & 0.04541    &  -0.00047  &  0.1    &(1)&53516.1400  &5515     & 0.3915   & 0.0955    &  0.00244  &   0.8   &(23)\\
28404.6215  &91       & 0.0013   & 0.03793    &  -0.00509 &  0.1    &(1)&54136.5301  &5649     & 0.4108   & 0.10658   &  -0.00261 &   0.8   &(24)\\
33756.429   &1247     & -0.0467  & -0.08096   &  0.03089  &   0.1   &(1)&54210.598   &5665     & 0.4046   & 0.0994    &  -0.01166 &   0.8   &(24)\\
33756.463   &1247     & -0.0127  & -0.04696   &  0.06489  &   0.1   &(1)&54187.48083 &5660     & 0.4356   & 0.13071   &  0.02022  &   0.8   &(25$^{*}$)\\
34455.498   &1398     & -0.0523  & -0.09582   &  0.03284  &   0.1   &(1)&54252.3029  &5674     & 0.4428   & 0.13705   &  0.02493  &   0.8   &(26)\\
\hline
\end{tabular}
\end{center}
\tiny
Note.  $^{*}$ the mean value of 3 times of light minimum\\
\tiny References: (1) \cite{hal80}; (2) \cite{zve33};
(3)\cite{kor34}; (4) \cite{zve37}; (5) \cite{die70}; (6)
\cite{die71a};
 (7) \cite{die71b}; (8) \cite{bbs73a}; (9) \cite{bbs73b}; (10) \cite{kiz74}; (11) \cite{bbs75}; (12) \cite{mar80};
 (13) \cite{isl84};  (14) \cite{poh82}; (15) \cite{tun81}; (16) \cite{bbs83}; (17) \cite{bbs85}; (18) \cite{isl92};
 (19) \cite{bla94}; (20) \cite{hub00}; (21) \cite{hub02}; (22) \cite{hub03}; (23) \cite{nag06}; (24) \cite{hub07};
 (25) \cite{bra07}; (26) The present author.
\end{table*}

\begin{figure}
\centering
\includegraphics[width=9.0cm]{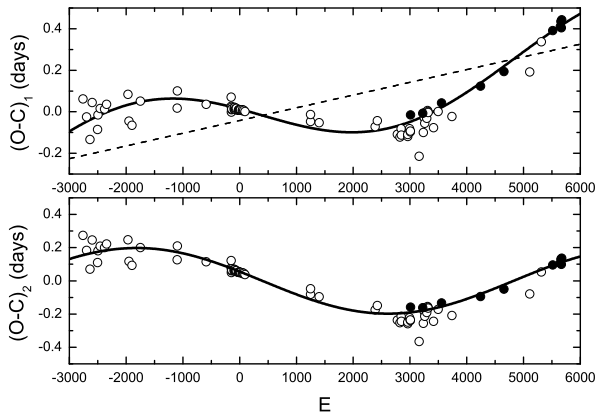}
\caption{$(O-C)$ diagram of WW Dra. The upper panel : $(O-C)_1$
diagram of WW Dra computed with Eq. (1). The open circles refer to
photographic or visual observations, filled circles refer to CCD or
photoelectric ones. The solid line refers to a combination of a
linear ephemeris and a cyclical period variation, and the dashed
line to a new linear ephemeris. The lower panel : $(O-C)_2$ curve of
WW Dra as described by the sinusoidal term(solid line), after
removing the linear term. The symbols are the same as in the upper
panel.} \label{Fig. 4}
\end{figure}

\begin{figure}
\centering
\includegraphics[width=9.0cm]{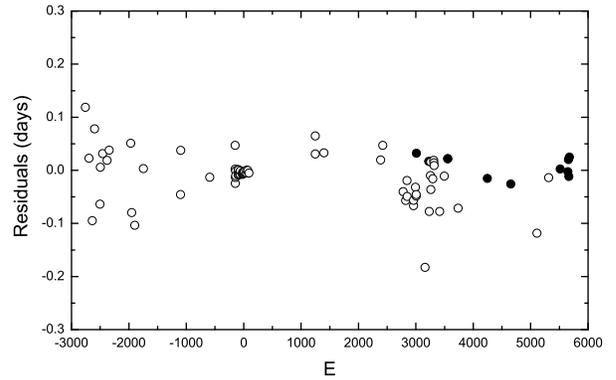}
\caption{The residuals \textbf{from fit with Eq. (2)}. The symbols
are the same as in the Fig. 4.} \label{Fig. 5}
\end{figure}

\section[]{Discussions and conclusions}

In Section 4, we displayed the existence of a cyclical period change
of WW Dra. This cyclical variation may be interpreted as due to
\textbf{either} the magnetic activity of one or both components
\cite[]{app92}, or by the LTTE via the presence of a tertiary
companion. With the following equation given by \cite{rov00},

\begin{eqnarray}
 \Delta P=\sqrt{2[1-\cos(2\pi P/P_{3})]} \times A ,
\end{eqnarray}
\noindent where $P_{3}$ is the period for the $(O-C)$ oscillation,
the rate of the period variation is \textbf{found to be} $\Delta P$/
$P$ = 5.29 $\times 10^{-7}$. In order to reproduce this cyclic
change, the required variation of the quadruple momentum $\Delta Q$
can be calculated with the following equation \cite[]{lan02},
\begin{eqnarray}
 \frac{{\Delta P}}{P}=-9\frac{\Delta Q}{Ma^{2}},
\end{eqnarray}
\noindent where \textit{a} is the separation between both components
\textbf{that can} be determined with the Kepler's third law,
\begin{eqnarray}
 M_{1}+M_{2}=0.0134\frac{a^3}{P^{2}},
\end{eqnarray}
\noindent where ${M_{1} = 1.36 M_{\odot}}$ and ${M_{2} = 1.34
M_{\odot}}$ \cite[]{alb99}. Combining Eq. (4) and Eq. (5), we
derived $\Delta Q_{1}$ = 2.04 $\times 10^{50}$ g $\textrm{cm}^{2}$
and $\Delta Q_{2}$ = 2.01 $\times 10^{50}$ g $\textrm{cm}^{2}$ for
both components, respectively. Assuming conservation of the orbital
angular momentum, \textbf{the total} $\Delta Q$ is on the order of
$10^{51}-10^{52}$ g $\textrm{cm}^{2}$ \cite[]{lan99}, \textbf{which
indicates} the values of $\Delta Q_{1}$ and $\Delta Q_{2}$ for WW
Dra are \textbf{not typical ones for the close binaries},
suggesting, that the mechanism of Applegate cannot interpret the
cyclical period variation of WW Dra. Moreover, the period of 112.2
\textbf{\textrm{yr}} for the $(O-C)$ oscillation of WW Dra is too
long in comparison with magnetic cycles in solar type single
\textbf{stars} and close binaries \cite[]{mac90, bia90}. Therefore,
the cyclical period change is more plausibly interpreted as
\textbf{due to} the presence of a third body.

The good sinusoidal fit shown in Fig. 4 suggests that the orbit of
the third body is approximately circular, which is different from
the result derived by \cite{alb99}. Using $a_{12}\textrm{sin}i' = A
\times c$, where $i'$ is the inclination of the orbit of the third
component and \textit{c} is the speed of light,
$a_{12}\textrm{sin}i'$ is computed to be 34.25($\pm1.66$) AU. Then
combining the following well-known equation,
\begin{equation}
f(m)
=\frac{4\pi^{2}}{G{P}_{3}^{2}}\times(a\prime_{12}\sin{i^{\prime}})^{3},
\end{equation}
\noindent with

\begin{equation}
f(m)=\frac{(M_{3}\sin{i_3^{\prime}})^{3}}
{(M_{1}+M_{2}+M_{3})^{2}}
\end{equation}
\noindent the mass function of the third body is computed to be
$f(m_3)= 3.19(\pm0.47)\,M_{\odot}$. In the formula, $M_{1}$,
$M_{2}$, and $M_{3}$ are the masses of the eclipsing pair and the
third companion, respectively, $G$ is the gravitational constant.
According to the same parameters (${M_{1} = 1.36 M_{\odot}}$ and
${M_{2} = 1.34 M_{\odot}}$) used by \cite{alb99}, the lowest mass of
the third body is calculated to be $M_{3}=6.43M_{\odot}$, and the
third body is orbiting the binary at a distance shorter than 14.4
AU. When the third body is coplanar to the eclipsing binary :
$i^{\prime} = i = 81.\,^{\circ}4$ (according to \cite{alb99}), its
mass is $M_{3}=6.57M_{\odot}$. Using the formula given by
\cite{may90},
\begin{equation}
 K_{RV}=\frac{2\pi}{P_{3}}\frac{a_{12}\textrm{sin}i_{3}}{\sqrt{1-e'^{2}}}
\end{equation}
\noindent where $K_{RV}$, $P_{3}$, $a_{12}$ are in kilometer per
second, years and AU, respectively, and considering the simplest
situation of $i_{3} = 90^{\circ}$, the semi-amplitude of the system
velocity accompanied by the light-time effect is approximately
calculated to be 9.09 $\textrm{km}~\textrm{s}^{-1}$, which is a
little less than the value determined by \cite{alb99}. According to
Allen's tables \cite[]{dri00}, the third companion is estimated to
be a $\sim$ B4 \textbf{star}. Therefore, it \textbf{could} be
discovered by spectroscopic observation. However, no spectral lines
of the third body were discovered up to now. It may be explained in
two possible ways: (1) the star was observed in the past in a
spectral range where the third body has no lines, or lines were
present but the poor resolution of available spectra did not allow
to detect them. Actually, it is difficult to find sufficient
spectral lines to determine radial velocity of B stars because their
rapid rotational velocity makes them too broad and weak to be
accurately measured, or (2) the third body is a candidate black hole
and it may play an important role in the evolution of this system.
The situation resembles that of V Pup \citep{qia08b}. More
observations are needed to check this hypothesis in the future. All
these make WW Dra a very interesting system to study.

\section*{Acknowledgments}

This work is partly supported by Chinese Natural Science Foundation (No.10973037, No.
10903026 and No.10778718), the National Key Fundamental Research Project through grant
2007CB815406, the Yunnan Natural Science Foundation (No. 2008CD157). We are indebted to the many
observers, amateur and professional, who obtained the wealth of
data on this eclipsing binary system listed in Table 5.

\bsp
\label{lastpage}
\end{document}